\documentclass[journal,fleqn,10pt,twocolumn,twoside,a4paper]{IEEEtran}
\usepackage{amsmath,amssymb}
\usepackage{graphicx}
\usepackage{cite}

\usepackage[linesnumbered,ruled]{algorithm2e} 
\SetKwInOut{Input}{Input}
\SetKwInOut{Output}{Output}

\usepackage{color}
\usepackage{multirow}
\usepackage{flushend} 

\begin{document}

\title{6G NTN Waveforms: A Comparison of OTFS, AFDM and OCDM in LEO Satellite Channels}

\author
{
\IEEEauthorblockN{
Baidyanath~Mandal\IEEEauthorrefmark{1}, 
Aniruddha~Chandra\IEEEauthorrefmark{1}, 
Rastislav~Róka\IEEEauthorrefmark{2}, \\
Jarosław~Wojtuń\IEEEauthorrefmark{3},
Jan~Kelner\IEEEauthorrefmark{3}, and
Cezary~Ziółkowski\IEEEauthorrefmark{3}\\
\vspace{5pt}
}
\IEEEauthorblockA{
\IEEEauthorrefmark{1}ECE Department, National Institute of Technology, Durgapur 713209, WB, India\\ 
\vspace{2pt}
}
\IEEEauthorblockA{
\IEEEauthorrefmark{2}Institute of Multimedia Information and Communication Technologies,\\ 
Slovak University of Technology, Bratislava 84104, Slovakia\\
\vspace{2pt}
}

\IEEEauthorblockA{
\IEEEauthorrefmark{3}Institute of Communications Systems, Faculty of Electronics, \\Military University of Technology, 00-908 Warsaw, Poland\\
\vspace{2pt}
}
Email: \{bm.24p10370, achandra.ece\}@nitdgp.ac.in, rastislav.roka@stuba.sk, \\ 
\{jaroslaw.wojtun, jan.kelner, cezary.ziolkowski\}@wat.edu.pl
}

\maketitle
\thispagestyle{empty} 

\begin{abstract}
Sixth generation (6G) physical layer (PHY) is evolving beyond the legacy orthogonal frequency division multiplexing (OFDM)-based waveforms. In this paper, we compare the bit error rate (BER) performance of three beyond-OFDM waveforms, namely, orthogonal time-frequency-space (OTFS) modulation, affine frequency division multiplexing (AFDM), and orthogonal chirp division multiplexing (OCDM), which are particularly suitable for the highly mobile non-terrestrial network (NTN) vertical of 6G. In order to characterize the effect of mobility and Doppler shift in low Earth orbit (LEO) satellites, we performed BER comparisons over four different NTN tapped-delay-line (TDL) models, TDL-A, TDL-B, TDL-C, and TDL-D, as specified in the 3rd generation partnership project (3GPP) technical report TR 38.811. After channel equalization, a minimum mean squared error with successive detection (MMSE-SD) algorithm was used to enhance the BER performance. It was found that AFDM and OTFS consistently outperformed OCDM across all TDL models, while AFDM performed better than OTFS in TDL-B and TDL-C, in the high signal-to-noise ratio (SNR) regime. The complete simulation framework is made available as an open-source code for quick validation and further development.
\end{abstract}


\section{Introduction} \label{sec:intro}
\begin{figure}[t]
    \centering
    \includegraphics[width=\columnwidth]{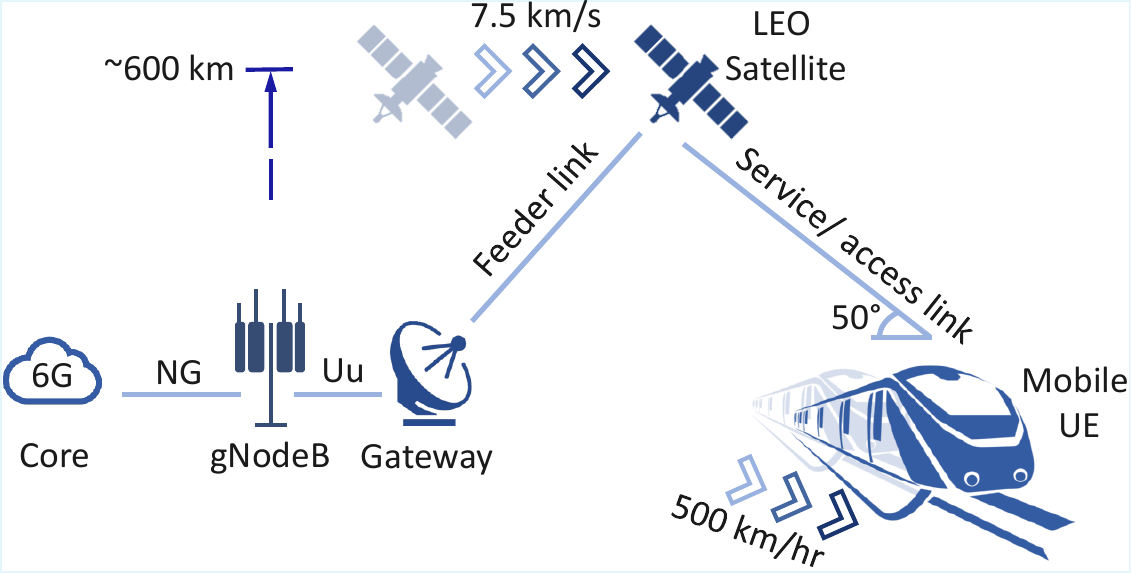}
    \caption{6G NTN architecture with a fast-moving LEO satellite and a highly mobile user equipment (UE). The gNodeB connects to the core via an NG interface, whereas it connects to the UE via a Uu interface through an NTN payload in a transparent manner.}
    \label{fig:usecase}
\end{figure}

Non-terrestrial networks (NTNs) have been an integral part of the 3rd generation partnership project (3GPP) releases from release 17 (R17) onwards for standardizing 5th generation (5G) new radio (NR) \cite{NTN6Gbook}. In 6G, whose study phase is marked by R20, NTNs are expected to play a key role in achieving ubiquity, transparency, reconfigurability, and resilience through NTN-Internet of Things (IoT) coalescence. There are a plethora of NTN nodes ranging from different types of satellites to high-altitude platforms (HAPs) and to unmanned aerial vehicles (UAVs). However, thanks to the recent agreements between commercial satellite operators and mobile network operators (e.g., Starlink T-Mobile direct handset to satellite service), low Earth orbit (LEO) satellites have become native to the cellular fabric. No wonder R20 is provisioning special study items for LEO satellites, including seamless handover between gNodeB and LEO satellite, Doppler/ high-velocity considerations, and synchronization without global navigation satellite systems (GNSS) \cite{jung2025sixth}. A typical NTN architecture with a LEO satellite and 3GPP interfaces \cite{lin20215g} for highly mobile scenarios is shown in Fig. \ref{fig:usecase}.     

\subsection{Motivation}
The relative motion between a UE and a LEO satellite results in a considerable Doppler shift. As specified in the 3GPP technical report TR~38.811, LEO satellite links operating in non-terrestrial networks may experience Doppler shifts as high as $\pm48$ kHz in the S-band ($2-4$ GHz), which can further increase to $\pm480$ kHz in the Ka-band ($26-40$ GHz) \cite{3GPP_TR_38_811}. Large Doppler-induced carrier frequency offsets (CFOs) lead to cyclic symbol misalignment and severe inter-carrier interference (ICI). The traditional orthogonal frequency division multiplexing (OFDM)-based waveforms, which worked well for terrestrial networks, are not optimal for such NTN scenarios. Orthogonal time-frequency-space (OTFS) modulation, affine frequency division multiplexing (AFDM), and orthogonal chirp division multiplexing (OCDM) are three major candidate waveforms introduced recently to overcome the limitations of OFDM. Thus, it is interesting to study their suitability in the context of an LEO satellite link.

\subsection{Literature Survey}
AFDM is a generalized multicarrier modulation introduced by Bemani \textit{et al.} \cite{bemani2023affine} that employs discrete affine Fourier transform (DAFT), which maps symbols into the twisted time-frequency domain, enabling flexible control over time–frequency localization and maintaining delay-Doppler orthogonality. OTFS modulation introduced by Hadani \textit{et al.} \cite{7925924otfs} shifts the domain to delay-Doppler, and makes the channel semi-static by reducing the effect of Doppler shifts. Raviteja \textit{et al.} \cite{8377159ravi2,8424569raviteja} significantly advanced OTFS by developing low-complexity message-passing–based detection and channel estimation algorithms that exploit delay--Doppler domain sparsity and developed input-output analysis of OTFS in delay-Doppler domain. To counter the doubly-dispersive nature, OCDM, which utilizes chirp spread spectrum, was proposed by Ouyang \textit{et al.} \cite{7523229ocdm}. The performance comparison of OTFS, AFDM, and OFDM  was shown by Rou \textit{et al.} \cite{rou2024orthogonal} for integrated sensing and communication (ISAC) applications. As far as comparisons over LEO satellite link is concerned, there is only one available study by Liu \textit{et al.} \cite{liu2025otfs}, where OTFS and OFDM are compared against each other.
\begin{figure*}[t]
    \centering
    \includegraphics[width=0.8\textwidth]{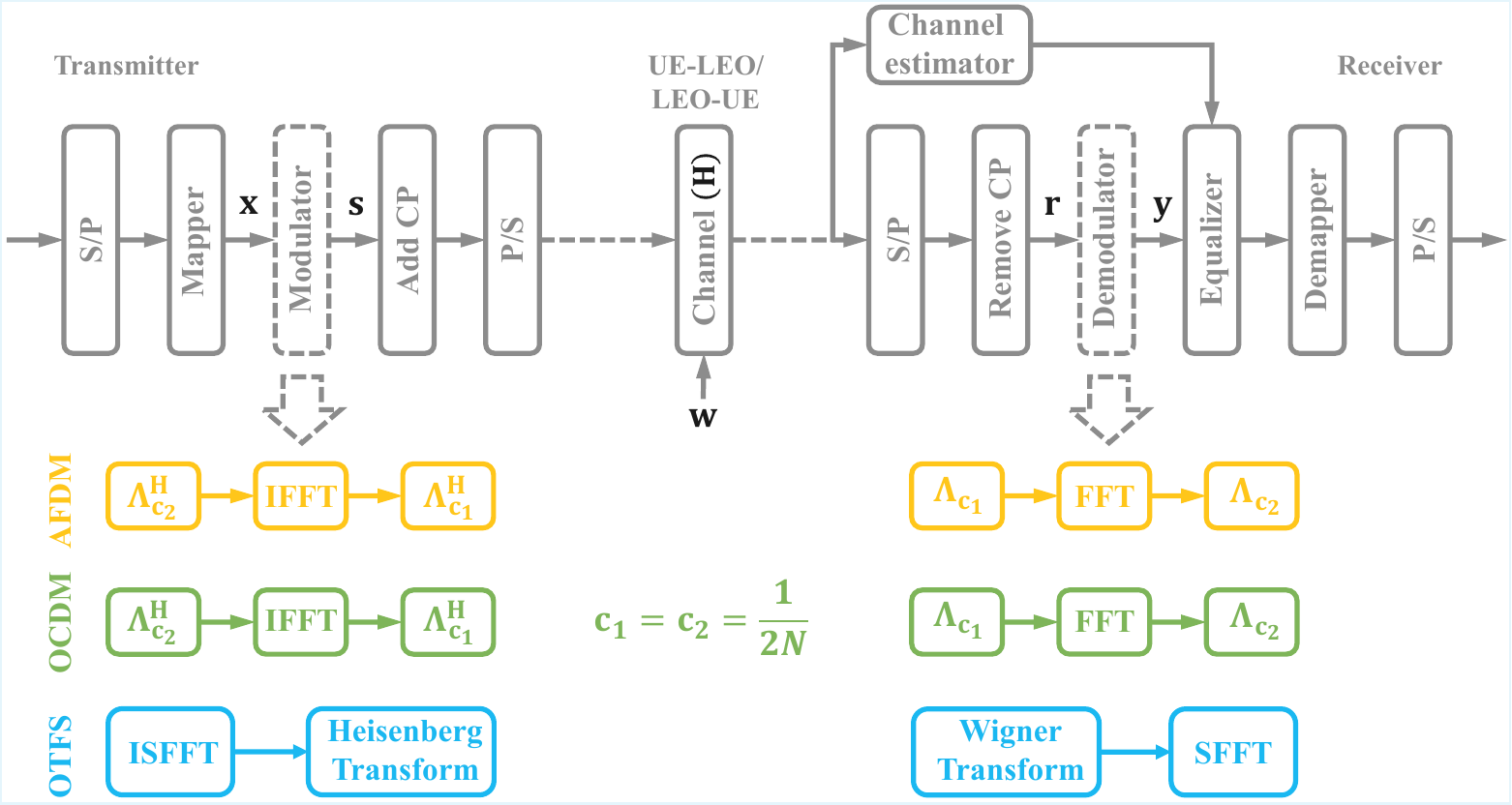}
    \caption{Transmitter and receiver schematic for waveform simulation. The modulator uses a discrete inverse transform, and the demodulator uses the corresponding discrete transform. OCDM $(c_1=c_2=1/(2N))$ and OFDM $(c_1=c_2=0)$ are special cases of AFDM. Channel estimator feeds a linear minimum mean square error equalizer.}
    \label{fig:wave}
\end{figure*}

\subsection{Contribution}
To address the gap, as mentioned in the previous subsection, we compare OTFS, AFDM, and OCDM waveform performance over LEO satellite links in this paper. Specifically, our contributions are twofold:
\begin{itemize}
\item First, we have built a MATLAB-based framework for simulating the bit error rate (BER) of OTFS, AFDM, and OCDM under identical system and channel conditions (e.g., channel type, SNR range, modulation type, and modulation order). A minimum mean squared error with successive detection (MMSE-SD) algorithm \cite{li2021otfs} was used after channel equalization, which resulted in an average SNR gain of $1-2$ dB. The source code is made available in an open GitHub repository \cite{GitHubRepo}.

\item Next, using the developed framework described above, we have compared BER of the three candidate waveforms in LTE satellite links. We have considered link parameters (e.g., Doppler shift, normalized delay and corresponding gains, Rician K-factor) as per the four tapped-delay-line (TDL) models, TDL-A, TDL-B, TDL-C, and TDL-D, conforming to 3GPP specifications \cite{3GPP_TR_38_811}. We have shown that OCDM is unusable as it hits an irreducible error floor. The performance of OTFS is comparable to AFDM in the low signal-to-noise ratio (SNR) regime, but for SNR $>15$ dB, AFDM starts outperforming OTFS.
\end{itemize}

\subsection{Organization}
The rest of the paper is organized as follows. Section \ref{sec:waveform_model} discusses the effective channel matrices required to simulate the waveforms, while Section \ref{sec: channel_model} describes NTN TDL models and MMSE-SD detection algorithm. Section \ref{sec: results} compiles the comparison results. Finally, the paper concludes in Section \ref{sec: conc}.

\section{Waveform Model} \label{sec:waveform_model}
OFDM performs well over semi-static TN channels, but in the case of a doubly dispersive channel, its performance degrades due to the presence of the Doppler effect \cite{1638663proakis}. Orthogonality between the subcarriers is lost, leading to inter-carrier interference (ICI). The energy associated with a subcarrier is concentrated at that particular frequency, and loss of orthogonality means spreading of the energy associated with that frequency.

Access and service links between a non-terrestrial LEO satellite and a terrestrial moving UE experience large relative motion. To mitigate the consequential Doppler effect, AFDM \cite{bemani2023affine} and OCDM \cite{7523229ocdm} have been proposed, in which chirp frequencies are employed for the subcarriers. These subcarriers are also orthogonal, but now their energy is not discretized at a particular frequency; the associated energy is spread in the time-frequency domain, making it very robust to counter the energy spreading due to the Doppler effect. OTFS \cite{7925924otfs,8424569raviteja,8503182chokalingam}, on the other hand, uses the delay-Doppler domain to counter the time selectivity and frequency selectivity of the channel.

Transmitter and receiver blocks for these three waveforms are shown in Fig.~\ref{fig:wave}. AFDM, OCDM, and OTFS are all derived from basic OFDM, where an inverse discrete transform is applied at the transmitter end, and its action is reversed with a direct discrete transform at the receiver end. 

In OFDM, the transmitter takes the input as a frequency domain signal vector $\mathbf{x} \in \mathbb{C}^{N \times 1}$, having $N$ complex symbols, and applies inverse fast Fourier transform (IFFT) to obtain a time domain signal vector,
\begin{equation}
    \mathbf{s}_{\mathrm{ofdm}} = \mathbf{F}_N^{\mathrm{H}} \mathbf{x},
\end{equation}
where $\mathbf{F}_N \in \mathbb{C}^{N \times N}$ is a $N$-point FFT matrix, and its conjugate transpose, $\mathbf{F}_N^{\text{H}}$, is a $N$-point IFFT matrix \cite{9100670Ramjee}. The noise corrupted version at receiver is given by, 
\begin{equation}
    \mathbf{r}_{\mathrm{ofdm}} = \mathbf{H} \mathbf{s}_{\mathrm{ofdm}} + \mathbf{w},
\end{equation}
where $\mathbf{H}$ and $\mathbf{w}$ denote channel matrix and noise matrix, respectively. The symbols recoverd at receiver end, $\mathbf{y}_{\mathrm{ofdm}} \in \mathbb{C}^{N \times 1}$, making use of $N$-point FFT is:
\begin{equation}
    \mathbf{y}_{\mathrm{ofdm}} = \mathbf{F}_N \mathbf{r}_{\text{ofdm}} = {(\mathbf{F}_N \mathbf{H} \mathbf{F}_N^{\mathrm{H}})} \mathbf{x} + \mathbf{F}_N \mathbf{w}.
\end{equation}
Thus, the effective channel matrix for OFDM is given by,
\begin{equation}
    \mathbf{H}_{\mathrm{eff}}^{\mathrm{ofdm}}=\mathbf{F}_N \mathbf{H} \mathbf{F}_N^{\mathrm{H}}.
\end{equation}
It is worth mentioning that the channel matrix operations follow circular convolution.

\subsection{AFDM}
In AFDM, the modulated signal vector, $\mathbf{s}_{\mathrm{afdm}} \triangleq \mathbf{IDAFT}\cdot\mathbf{x}$; $\mathbf{s}_{\text{afdm}} \in \mathbb{C}^{N\times 1}$, is obtained following an inverse discrete affine Fourier transform (IDAFT) \cite{bemani2023affine,rou2025affine}:
\begin{equation}
    \mathbf{s}_{\mathrm{afdm}} 
    = (\mathbf{\Lambda}_{c_1} \mathbf{F}_N \mathbf{\Lambda}_{c_2})^{\mathrm{H}} \mathbf{x} 
    = \mathbf{\Lambda}_{c_2}^{\mathrm{H}} \mathbf{F}_N^{\mathrm{H}}    \mathbf{\Lambda}_{c_1}^{\mathrm{H}} \mathbf{x},
\end{equation}
where, the diagonal chirp matrix is defined as $\mathbf{\Lambda}_{c_i} \triangleq \mathrm{diag}\left[\exp(-j 2\pi c_i n^2) \right]_{n=0}^{N-1}; \mathbf{\Lambda}_{c_i} \in \mathbb{C}^{N \times N}$, where $c_{1,2}$ are two controllable digital chirp rates tuned in such a way that multipaths are resolvable. AFDM is capable of achieving complete orthogonality within the integer-normalised delay-Doppler domain under the condition, $2(f_{\max} + \xi)(\ell_{\max} + 1) + \ell_{\max} \leq N$, where $f_{\max}$ and $\ell_{\max}$ represent the maximum values for normalised digital Doppler shift and channel delay, respectively, while the term, $\xi \in \mathbb{N}_0$, specifies the number of extra guard components surrounding the diagonals to mitigate potential Doppler-domain interference. When this orthogonality requirement is satisfied, the ideal chirp frequencies for the AFDM waveform are determined by:
\begin{equation}
c_1 = \frac{2(f_{\max} + \xi) + 1}{2N}, \quad \text{and} \quad c_2 \ll \frac{1}{N},
\end{equation}

The received signal after passing through the UE-LEO/ LEO-UE channel is given by, $\mathbf{r}_{\mathrm{afdm}} = \mathbf{H} \cdot \mathbf{s}_{\mathrm{afdm}}  +\mathbf{w}$. Next, $N$-point DAFT operation is performed on the received signal, to obtain $\mathbf{y}_{\mathrm{afdm}} \triangleq \mathbf{DAFT} \cdot \mathbf{r}$, ; $\mathbf{y}_{\mathrm{afdm}} \in \mathbb{C}^{N\times 1}$, i.e., 
\begin{equation}
    \mathbf{y}_{\mathrm{afdm}} = \mathbf{DAFT} \cdot (\mathbf{H} \mathbf{s}_{\text{afdm}} + \mathbf{w}) = \mathbf{H}_{\mathrm{eff}}^{\text{afdm}} \mathbf{x} + \mathbf{DAFT} \cdot \mathbf{w},
\end{equation}
where, $\mathbf{DAFT} \triangleq \mathbf{\Lambda}_{c_1} \mathbf{F}_N \mathbf{\Lambda}_{c_2} $, whereas $\mathbf{H}$ and $\mathbf{H}_{\mathrm{eff}}^{\text{afdm}}$ are the doubly-dispersive channel matrix and effective channel matrix for AFDM, respectively.

\subsection{OCDM}
OCDM uses inverse discrete Fresnel transform (IDFnT) in the transmitter section and the discrete Fresnel transform (DFnT) in the receiver section. OCDM may be studied as a special case of  AFDM  with chirp parameters $c_1=c_2=1/(2N)$. Fig.~\ref{fig:wave} describes the reduction of an AFDM transreceiver to OCDM \cite{rou2024orthogonal}.

\subsection{OTFS}
In OTFS, the signal is placed in the delay-Doppler domain (as opposed to the time-frequency domain for AFDM/ OCDM), and then converted to the time domain. In time-frequency analysis, each path has different frequency fluctuations at different time instants, but in the delay-Doppler domain, each path has only two components: the delay of that path and the maximum Doppler associated with that path. This makes the channel sparse, and hence signal recovery can make use of this sparsity \cite{otfsBook, 9392379saif}.

For OTFS modulation, an inverse symplectic fast Fourier transform (ISFFT) is executed first to convert the delay-Doppler domain to the time-frequency domain, followed by a Heisenberg transform to convert the signal from the time-frequency domain to the time domain. The demodulation consists of a Wigner transform followed by SFFT. The information symbols in OTFS  are arranged in a two-dimensional matrix,
$\mathbf{x}\in\mathbb{C}^{K\times L}, N = KL$, and after modulation the final OTFS signal vector, $\mathbf{s}_{\mathrm{otfs}} \in \mathbb{C}^{KL \times 1}$, can be expressed as: 
\begin{align}
\mathbf{s}_{\mathrm{otfs}} 
& = \operatorname{vec} \left(\mathbf{P}^{\mathrm{tx}} \mathbf{F}_K^{-1} 
\left(\mathbf{F}_K \mathbf{x} \mathbf{F}_L^{-1} \right)   \right)
\nonumber\\
& = \left( \mathbf{F}_L^{-1} \otimes \mathbf{P}^{\mathrm{tx}} \right) \operatorname{vec}(\mathbf{x}),
\end{align}
where $\mathbf{F}_K$ and $\mathbf{F}_L$ are part of ISFFT and represent $K$-point and $L$-point FFT matrices, the diagonal matrix $\mathbf{P}^{\mathrm{tx}}$ is a transmit
pulse-shaping filter realizing Heisenberg transform, the operator $\operatorname{vec}(\cdot)$ stacks the columns of a matrix into a single column vector, while $\otimes$ denotes the Kronecker product. The received  signal, $\mathbf{y}_{\mathrm{otfs}} \in \mathbb{C}^{KL \times 1}$ is given by 
\begin{align}
\mathbf{y}_{\mathrm{otfs}} & =
\left( \mathbf{F}_L \otimes \mathbf{P}^{\mathrm{rx}} \right)
\left( \mathbf{H} \mathbf{s}_{\mathrm{otfs}} + \mathbf{w} \right) 
\nonumber\\
&=
\mathbf{H}_{\mathrm{eff}}^{\mathrm{otfs}} \mathbf{x}
+
\left( \mathbf{F}_L \otimes \mathbf{P}^{\mathrm{rx}} \right)\mathbf{w},
\end{align}
where, $\mathbf{P}^{\mathrm{rx}}$ is the receive pulse-shaping filter realizing Wigner transform and $\mathbf{H}_{\mathrm{eff}}^{\mathrm{otfs}}$ is the effective channel matrix.


\section{Channel Model and Detection}\label{sec: channel_model}

\subsection{Tapped Delay Line Models}
The channel adopted for simulation is a doubly dispersive linear time variant (LTV) channel which is quasi-static, i.e. during the transmission of the signal, the relative velocity of scatterers, number of scatterers, and other physical characteristics of the channel remain the same. For an LTV channel having $M$ number of independent and resolvable paths, where path gain coefficient, Doppler frequency, and time delay associated with the $m$-th path is given by $g_m$, $\nu_m$, and $\tau_m$, respectively, the channel impulse response $h(t,\tau)$ in the time-delay domain is represented as $h(t,\tau) = \sum_{m=1}^{M} g_m \exp(- j 2\pi \nu_m t) \delta(t-\tau_m)$. The baseband received signal, $y(t)$, is a convoluted version of the input signal $s(t)$, given by $y(t)= \int_{-\infty}^{\infty} s(t-\tau)h(t,\tau) \mathrm{d} \tau$, and using the definition of $h(t,\tau)$, it can be further expressed as, $y(t)= \sum_{m=1}^{M} g_m \exp(- j 2\pi \nu_m t) s(t-\tau_m)$. Quite evidently, an LTV channel can be tabulated in the form of a tapped-delay-line (TDL) model where multiple delayed versions of the input signal are multiplied with an associated tap gain.

Four TDL models for NTN are specified in 3GPP standard TR 38.811 \cite{3GPP_TR_38_811}, where the first two models, TDL-A and TDL-B, represent non-line-of-sight (NLOS) channel power delay profile (PDP), whereas the rest two, TDL-C and TDL-D, are constructed for characterizing line-of-sight (LOS) cases. All these four TDL models are summarised in Table \ref{tab:pdp_leo}. It may be noted that all delays mentioned in Table \ref{tab:pdp_leo} are normalized and should be multiplied with the target root mean square (RMS) delay spread before simulation.

\begin{table}[ht]
\centering
\caption{Power delay profile (PDP) of LEO satellite channel models}
\label{tab:pdp_leo}
\begin{tabular}{|c|c|c|c|c|c|}
\hline
&&&&&\\[0.1mm]
\textbf{Model} & \textbf{Tap \#} & \textbf{Delay} & \textbf{Power [dB]} & \textbf{Fading} & \textbf{$K$-Factor} \\[1.5mm]
\hline
&&&&&\\[0.1mm]
\multirow{3}{9mm}{TDL-A} 
& 1 & 0      & 0      & Rayleigh & -- \\
& 2 & 1.0811 & -4.675 & Rayleigh & -- \\
& 3 & 2.8416 & -6.482 & Rayleigh & -- \\[2mm]
\hline
&&&&&\\[0.1mm]
\multirow{4}{9mm}{TDL-B} 
& 1 & 0      & 0       & Rayleigh & -- \\ 
& 2 & 0.7249 & -1.973  & Rayleigh & -- \\
& 3 & 0.7410 & -4.332  & Rayleigh & -- \\
& 4 & 5.7392 & -11.914 & Rayleigh & -- \\[2mm]
\hline
&&&&&\\[0.1mm]
\multirow{3}{9mm}{TDL-C} 
& \multirow{2}{1.2mm}{1} 
    & 0       & -0.394  & LOS      & 10.224 dB \\
&   & 0       & -10.618 & Rayleigh & -- \\
& 2 & 14.8124 & -23.373 & Rayleigh & -- \\[2mm]
\hline
&&&&&\\[0.1mm]
\multirow{4}{9mm}{TDL-D} 
& \multirow{2}{1.2mm}{1} 
    & 0      & -0.284  & LOS      & 11.707 dB \\
&   & 0      & -11.991 & Rayleigh & -- \\
& 2 & 0.5596 & -9.887  & Rayleigh & -- \\
& 3 & 7.3340 & -16.771 & Rayleigh & -- \\[2mm]
\hline
\end{tabular}
\end{table}
While the model recommends Jake’s spectrum for calculating Doppler shift due to UE mobility, it gives the following expression for calculating the additional Doppler shift due to the satellite motion
\begin{equation}
\alpha_{addl} = \left( \frac{v_{\mathrm{sat}}}{c} \right)
\left( \frac{R}{R+h}\right) \cos (\phi_{\mathrm{elev}}) f_c ,
\end{equation}
where $v_{\mathrm{sat}}$ denotes satellite speed, $c$ denotes the speed of light,
$R$ denotes the Earth radius, $h$ denotes the satellite altitude,
$\phi_{\mathrm{elev}}$ denotes the satellite elevation angle, and
$f_c$ denotes the carrier frequency.

\subsection{Detection Algorithm}
Using an equalizer, the estimated transmit symbol vector $\tilde{\mathbf{x}}$ is obtained by linearly filtering the received vector $\mathbf{y}$ with the linear minimum mean square error (LMMSE) weight matrix, $\tilde{\mathbf{x}} = \mathbf{W}_{\mathrm{LMMSE}} \cdot \mathbf{y}$, where the weight matrix is given by,
\begin{equation}
    \mathbf{W}_{\mathrm{LMMSE}} =
    \left( \mathbf{H_{eff}}^H \mathbf{H_{eff}} + \sigma_n^2 \mathbf{I} \right)^{-1} \mathbf{H_{eff}}^H,
\end{equation}
$\mathbf{H_{eff}}$ is the effective channel matrix for the three waveform models considered, and $\sigma_n^2$ is the noise variance. 

Unlike conventional LMMSE detection, which treats all symbols equally, minimum mean square with successive detection (MMSE-SD) \cite{li2021otfs} leverages the post-detection signal-to-interference-noise ratio (SINR) to prioritize symbol detection as presented in Algorithm~\ref{alg:mmse_sd}. The SINR for $k^{th}$~undetected symbol is expressed as 
\begin{equation}
    \mathrm{SINR}_k = \frac{|\mathbf{w}_k \mathbf{h}_k|^2}{\sum_{j\neq k} |\mathbf{w}_k \mathbf{h}_j|^2 + \sigma_n^2 \|\mathbf{w}_k\|^2},
\end{equation}
where $\mathbf{w}_k$ is $k$-th row of $\mathbf{W}_{\mathrm{LMMSE}}$, $\mathbf{h}_j$ is the $j$-th column of $\mathrm{\mathbf{H_{eff}} }$, and $\mathrm{|\cdot|}$ and $||\cdot||$ have usual of connotation of absoluteness and second norm. The symbols are arranged in the order of decreasing SINR to choose the index $q_i$. The symbol $ \mathrm{\hat{x}}_{q_i}$ is estimated and hard decision are applied on  it i.e $\mathbf{\tilde{x}=Q(\hat{x})}$ and successive interference cancellation  is performed by removing the  detected symbol~$ \tilde{x}$ from the received vector $y^{i}$ and equating the $q_i$-th column of $\mathbf{H_{eff}}$, thereby improving the reliability of subsequent detections. 

\begin{algorithm}[ht]
\label{alg:mmse_sd}
\DontPrintSemicolon
\Input{Received vector $\mathbf{y}$, channel matrix $\mathbf{H}_{\text{eff}}$,
noise variance $\sigma_n^2$}
\Output{Detected symbol vector $\tilde{\mathbf{x}}$}

Initialize $i=1$, $\mathbf{y}^{(1)}=\mathbf{y}$, $\mathbf{H}^{(1)}=\mathbf{H}_{\text{eff}}$ \;

\For{$i \gets 1$ to $N$}
{
$\mathbf{W}^{(i)} :=
(\mathbf{H}^{(i)H}\mathbf{H}^{(i)}+\sigma_n^2\mathbf{I})^{-1}
\mathbf{H}^{(i)H}$\;
$\mathrm{SINR}_k^{(i)} := \displaystyle \frac{|\mathbf{w}_k \mathbf{h}_k|^2}{\sum_{j\neq k} |\mathbf{w}_k \mathbf{h}_j|^2 + \sigma_n^2 \|\mathbf{w}_k\|^2}$\;
$q_i := \arg\max_k \text{SINR}_k^{(i)}$\;
$\hat{x}_{q_i} := \mathbf{w}_{q_i}\mathbf{y}^{(i)}$\;
$\tilde{x}_{q_i} := \ {Q}(\hat{x}_{q_i})$\;
$\mathbf{y}^{(i+1)} := \mathbf{y}^{(i)}-\tilde{x}_{q_i}(\mathbf{H}^{(i)})_{q_i}$\;
$(\mathbf{H}^{(i+1)})_{q_i} := \mathbf{0}$\;
}

\Return $\tilde{\mathbf{x}}$\;
\caption{MMSE-SD Iterative Detection}
\end{algorithm}


\section{Simulation Results and Discussions} \label{sec: results}

\subsection{Simulation Parameters}
\begin{figure}[t]
    \centering
    \includegraphics[width=0.9\columnwidth]{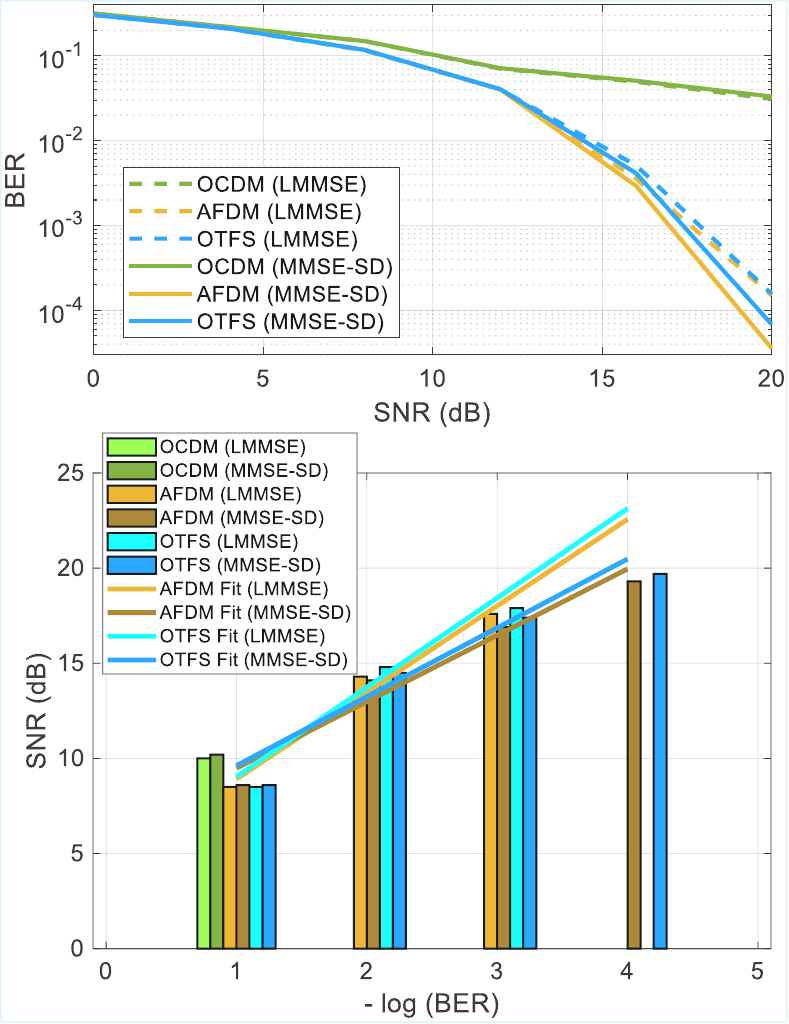}
    \caption{Comparison of different waveforms over NTN TDL-C channel with LMMSE and MMSE-SD: [Top] BER performance comparison, and [Bottom] Comparison of SNR requirement for achieving a target BER.}
    \label{fig:Comp}
\end{figure}

\begin{figure}[t]
    \centering
    \includegraphics[width=\columnwidth]{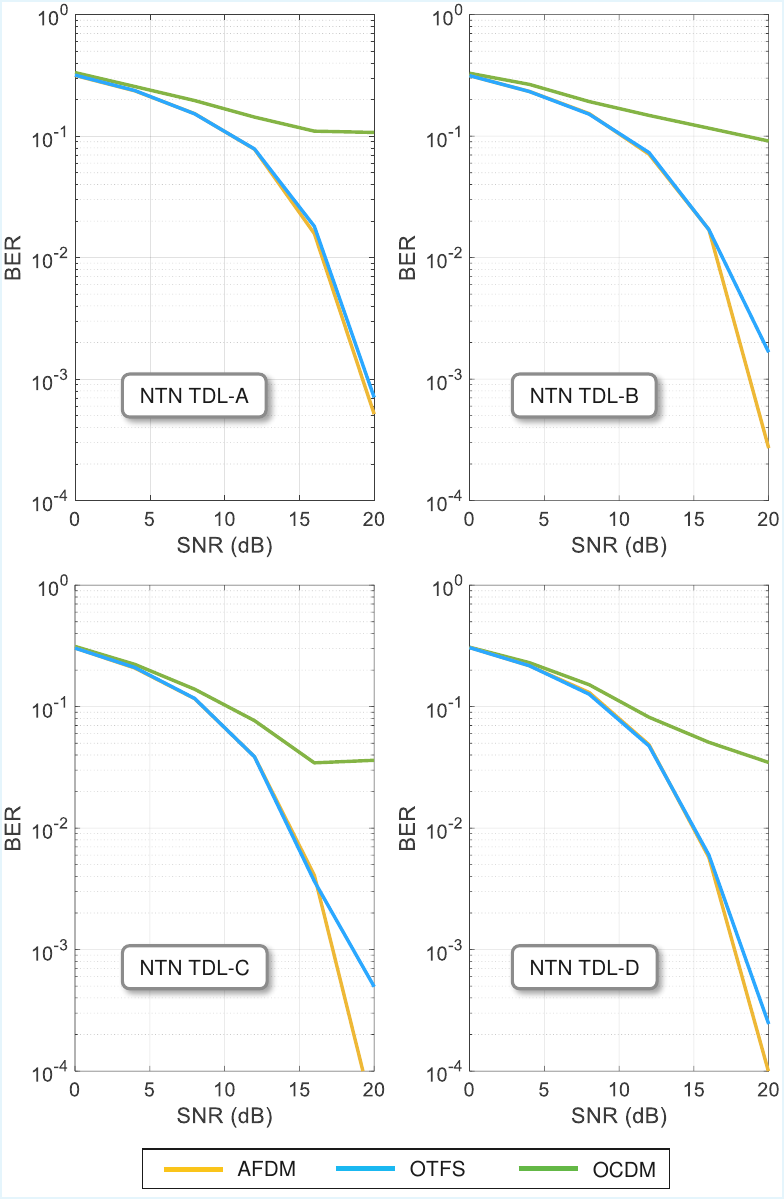}
    \caption{BER performance comparison of AFDM, OCDM and OTFS with MMSE-SD over different NTN TDL channels.}
    \label{fig:BER}
\end{figure}

In the simulations, integer-valued delay taps and fractional Doppler taps are assumed to dynamically adjust the grid size. The channel gains are modelled as random Gaussian variables, where each complex channel coefficient is generated with zero mean and a variance equal to the inverse of the number of propagation paths. For each channel realisation, a Jake's Doppler spectrum is assumed. The Doppler shifts are time-varying, and the Doppler shift associated with the $i$-th propagation path is given by, $\alpha_i = \alpha_{\max} \cos(\theta_i)$, where $\theta_i$ is uniformly distributed over the interval $[-\pi, \pi]$, and $\alpha_{\max}$ denotes the maximum Doppler shift.

For all three waveforms, a carrier frequency of $2.55$ GHz and a subcarrier spacing of $15$ kHz are considered. Further, quadrature amplitude modulation (QAM) with a modulation order of 16 is used for symbol mapping. For AFDM, the frame length is chosen as $N = 256$, while for OTFS, the frame is configured with $K = 16$ and $L = 16$. The cyclic prefix was suppressed for BER comparison.

For the LEO satellite model, the altitude is set to $600$ km, the satellite velocity is assumed to be $7.5$ km/s, and the elevation angle is fixed at $50^\circ$. It is assumed that the UE is moving at $500$ km/h, emulating a passenger in a next-generation high-speed train. The maximum Doppler shift with this setting is approximately $\alpha_{\max}=491$ Hz.

\subsection{Comparison of Detection Algorithms}
Fig. \ref{fig:Comp} demonstrates the comparison between basic LMMSE and MMSE-SD algorithms. The BER comparison shows that there is little effect of MMSE-SD on OCDM, but it significantly improves AFDM and OTFS performance. The same is more prominent from the bar graph below, where the fit lines for both AFDM and OTFS show a much lower slope with MMSE-SD, denoting that the SNR requirement to achieve a stringent target BER increases at a lesser rate.

\subsection{BER Performance Comparison}
Finally, in Fig. \ref{fig:BER}, we present a detailed comparison of BER performance for all three candidate waveforms over the four considered NTN-TDL channel models.

Looking at the top-left figure, it is evident that the OCDM performance starts to deviate from AFDM and OTFS at around an SNR of $5$ dB and hits an irreducible error floor after $15$ dB. AFDM and OTFS perform similarly in terms of BER for the NTN TDL-A LEO channel. This result also follows for the LEO NTN TDL-B channel till $17$ dB, but afterwards it starts performing better than OTFS.

NTN TDL-C has strong LOS components but with only three paths; in this scenario, the BER of all the waveforms improves, even OCDM exhibits a BER lower than the NLOS case. AFDM again starts performing better as the BER of AFDM falls with a higher gradient than that of OTFS BER. In the NTN TDL-D type channel with LOS components and four paths, the BER performance of AFDM and OTFS are almost similar, and the difference in slope is observable after $18$ dB. 

Overall, AFDM performs best in terms of the BER performance, and it achieves better diversity as the number of paths increase. From Fig. \ref{fig:BER} one can observe that AFDM can achieve a BER of $10^{-4}$ within $20$ dB SNR for LOS situations and on average results in an SNR gain of about $2$ dB.

\section{Conclusion}\label{sec: conc}
A comparison of OTFS, AFDM, and OCDM over a satellite link between a terrestrial UE and an LEO satellite is presented in this paper. It was found that at an elevation angle of $50^\circ$, the BER of AFDM is lowest among the three waveforms, while OTFS performs close to AFDM. The BER performance of OCDM is worst among the three. The trend is consistent over both LOS and NLOS channel models provided by 3GPP specifications.

\section*{Acknowledgment}
This work was a part of research activities within the framework of the National Scholarship Programme of the Slovak Republic administered by SAIA, Bratislava. This work was co-funded by the National Science Centre, Poland, under the OPUS call in the Weave program, under research project no. 2021/43/I/ST7/03294 acronym `MubaMilWave’, and chip-to-startup (C2S) program no. EE-9/2/2021-R\&D-E sponsored by MeitY, Government of India. 

\bibliographystyle{./IEEEtran}    
\bibliography{./myBib}

\end{document}